\documentclass{article}

\PassOptionsToPackage{numbers,compress}{plainnat}
\usepackage[final]{airov26}
\usepackage[utf8]{inputenc} % allow utf-8 input
\usepackage[T1]{fontenc}    % use 8-bit T1 fonts
\usepackage{booktabs}       % professional-quality tables
\usepackage{amsfonts}       % blackboard math symbols
\usepackage{nicefrac}       % compact symbols for 1/2, etc.
\usepackage{microtype}      % microtypography
\usepackage{xcolor}         % colors

\usepackage{amsmath}
\usepackage{cleveref}
\usepackage{graphicx}
\usepackage{doi}
\usepackage{url}
\usepackage{hyperref}

\title{Evaluation of Anatomical Shape Priors in Deep Learning-Based Cardiac Multi-Compartment Segmentation}

\author{%
  Michael Hudler, Franz Thaler, Martin Urschler\\
  Institute for Medical Informatics, Statistics and Documentation\\
  Medical University of Graz\\
  \texttt{\{michael.hudler, martin.urschler\}@medunigraz.at} \\
}

\begin{document}

\maketitle

\begin{abstract}
Whole-heart multi-compartment CT segmentation is clinically important, but standard CNNs do not explicitly enforce anatomical plausibility. Based on statistics derived from the training data, we evaluate whether lightweight explicit shape priors, implemented as shape-aware losses and spatial label distribution heatmap-guided U-Net variants, improve 3D cardiac segmentation on MM-WHS CT and WHS++. Across all experiments, a standard 3D U-Net surprisingly remained a very strong baseline, with handcrafted priors yielding at best marginal and inconsistent changes and often degrading performance. These results suggest that the baseline already captures substantial implicit anatomical regularities and that future gains will likely require more expressive learned priors rather than simple handcrafted anatomical shape constraints.
\end{abstract}

\section{Introduction}
Multi-compartment whole-heart segmentation from cardiac CT~\cite{Zhuang2019-uj} is a core task in medical image analysis because it supports quantitative assessment of clinical parameters like ejection fraction, builds the foundation for treatment planning or simulation~\cite{Zappon2025-dc}, and enables image-guided interventions. Deep learning~\cite{LeCun2015-ji}, especially 3D U-Net variants~\cite{Ronneberger2015-ih,Cicek2016-ty}, has become the dominant approach for this problem due to its strong multiscale feature extraction capabilities and its support for accurate localization~\cite{Payer2018-ma}. 
Their success is explained by the clever combination of encoder-decoder feature extraction, multiscale context modeling, and skip connections that preserve spatial detail.
However, these models are primarily appearance-driven and do not explicitly encode anatomical shape knowledge as was prominently done in the pre-deep learning era via statistical shape models~\cite{Cootes1995-gb, Heimann2009-yi}. This gap motivates the study of shape priors in deep learning-based segmentation~\cite{Bohlender2023-mi}.

This work evaluates whether explicit shape priors improve whole-heart multi-compartment CT segmentation beyond a strong 3D U-Net baseline. Rather than tediously building a full statistical shape-modeling pipeline, our study tests lightweight priors that can be incorporated directly into the training objective or 3D network design. The central question is whether such priors provide measurable benefit on modern deep learning baselines for seven-class cardiac CT segmentation, involving ventricles, both atria, myocardium, and the great vessels. The main finding is negative but clear: in the studied setting, explicitly designed shape priors did not consistently improve performance over a well-trained 3D U-Net baseline.

\section{Methods}
\textbf{Dataset and preprocessing.} Experiments are based primarily on the CT subset of the MM-WHS challenge dataset~\cite{Zhuang2019-uj}, comprising seven foreground classes: left ventricle, right ventricle, left atrium, right atrium, myocardium, ascending aorta, and pulmonary artery. The benchmark provides 20 annotated CT scans for training and 40 CT test scans evaluated with the official hidden-label evaluation script used for the Challenge. To complement this with accessible ground truth, evaluation was extended using the second half of the WHS++ training set (20 CT cases), which corresponds to the publicly released extension of the MM-WHS CT dataset. Images and labels were reoriented to a common anatomical convention, resampled isotropically, centered using label centroids, and embedded into a standardized field of view. A Procrustes-based alignment~\cite{Schoenemann1966} of training labels was further used to derive population heatmaps representing average spatial label distributions in the registered space.

\textbf{Baseline model.} The reference model is a standard 3D U-Net~\cite{Ronneberger2015-ih, Cicek2016-ty} with single-channel CT input and eight output classes including background. It uses a conventional encoder-decoder design with skip connections with 64 base channels and doubling the number of channels at each downsampling step. It uses LeakyReLU activations~\cite{Maas2013-ak} and is trained with a combined Generalized Dice~\cite{Sudre2017-cp} and Cross-Entropy loss. This baseline serves as the main point of comparison throughout the study.

\textbf{Shape-aware losses.} Three families of explicit regularizers were evaluated in combination with the baseline loss. \textit{Volume regularization} penalizes deviations from expected compartment volumes estimated from the training set via the label-specific volume means and standard deviations. Moment-based \textit{shape regularization} compares soft first- and second-order spatial moments of predictions to reference shape moments (centroids, ellipsoids) from the training set via L2 distance. \textit{Anatomical relation} loss constrains pairwise distances and angular relations between class centroids via reference angle statistics derived from the training data. All losses aim to inject coarse anatomical prior knowledge without changing the overall segmentation backbone.

\begin{figure}[t]
\centering
\includegraphics[width=0.75\textwidth]{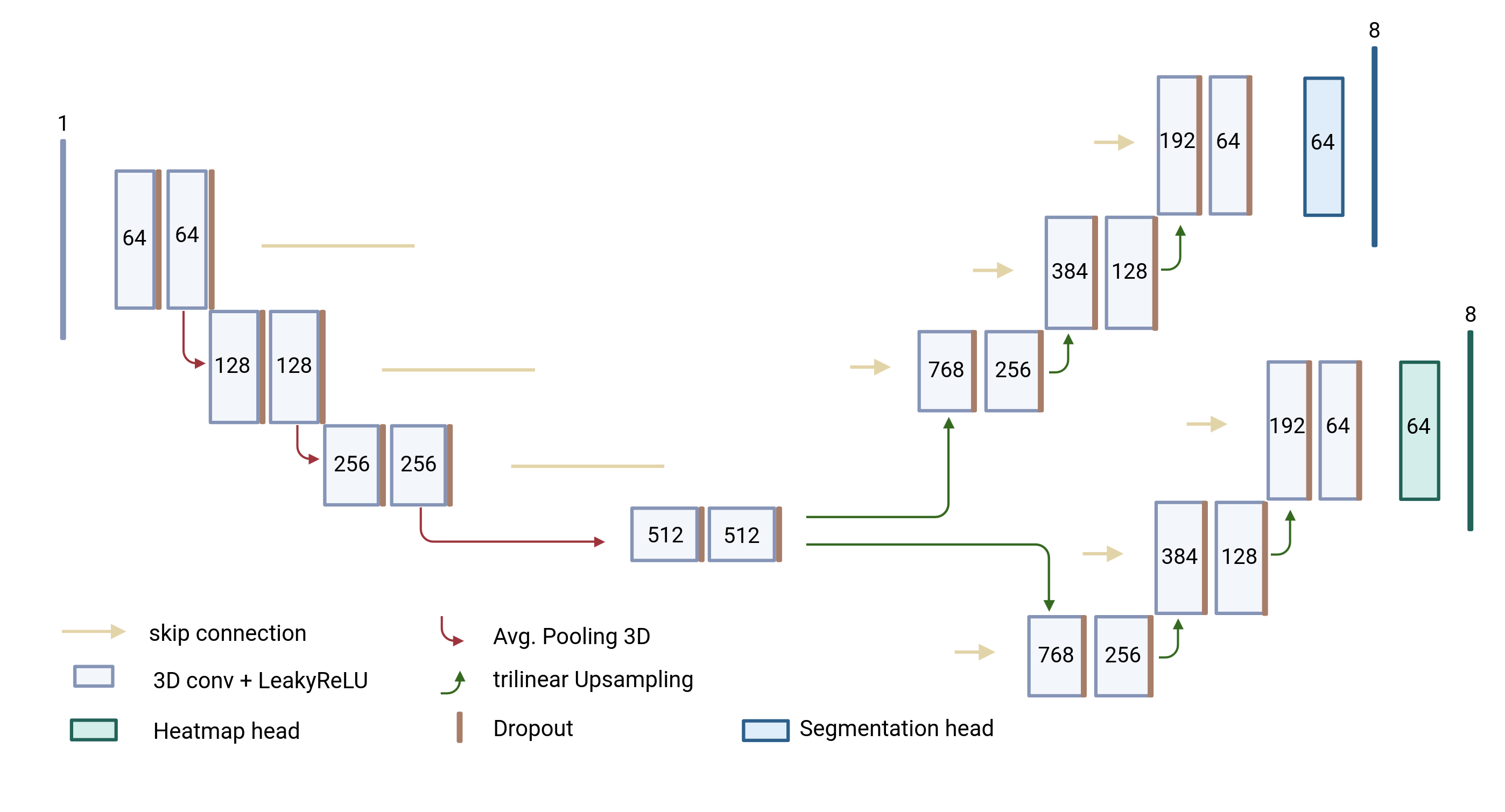}
\caption{Exemplary architecture of one of our proposed networks incorporating label distribution heatmaps: The 2-Decoder network, with separate decoders for label segmentation and heatmap predictions, thus forcing the encoder to extract features supporting both predictions.} \label{fig:2decoderarchitecture}
\end{figure}

\textbf{Architectural priors.} In addition to loss-based priors, we also investigated population-level multiclass probability heatmaps derived from aligned labels of the training dataset. These shape priors were integrated into several U-Net variants: a model with an auxiliary heatmap prediction head attached to the last decoder layer, a multilayer deep-supervision version of the latter architecture (\textit{HM multilayer}), a two-decoder network with separate segmentation and heatmap branches (\textit{2-Decoder}, see exemplary architecture depicted in Fig.~\ref{fig:2decoderarchitecture}), a dual-encoder network that processes image and heatmap inputs in parallel (\textit{2-Encoder}), and a cascaded three-U-Net architecture for coarse prediction and refinement (\textit{Cascaded}).

\textbf{Experimental setup.} Models were trained on cropped regions of interest at \(64^3\) and \(128^3\) input resolution using the same extensive geometric and intensity data augmentation for all architectures. Evaluation used Dice, Jaccard, Hausdorff distance (HD), and Average Symmetric Surface Distance (ASSD), as overlap- and boundary-based metrics, respectively. Qualitative comparisons were additionally assessed on WHS++, since the MM-WHS test set did not provide ground truth segmentations.

\section{Results}
Table~\ref{tab:results} summarizes the main findings. More results, additional descriptions of methods and implementation details can be found in~\cite{Hudler2026}. On MM-WHS at \(64^3\), the baseline achieved \(90.85\%\) Dice, \(83.63\%\) Jaccard, \(7.64\) mm HD, and \(1.03\) mm ASSD. Volume and moment regularization were essentially tied with the baseline (\(90.85\%\) and \(90.84\%\) Dice), whereas the anatomical relation loss reduced performance to \(88.98\%\) Dice. Thus, simple handcrafted losses did not harm but also did not improve the already strong baseline.

\begin{table}[h]
\caption{Main quantitative results. MM-WHS values report Dice, Jaccard, HD, and ASSD; WHS++ reports Dice only, as summarized in the thesis. Best values per block are in bold.}
\label{tab:results}
\centering
\small
\setlength{\tabcolsep}{4pt}
\begin{tabular}{llcccc}
\toprule
\textbf{Setting} &  & Dice (\%) & Jaccard (\%) & HD (mm) & ASSD (mm) \\
& Method & & & & \\
\midrule
\multicolumn{6}{l}{\textbf{MM-WHS CT, \(64^3\), shape-aware losses}} \\
& Baseline & \textbf{90.85} & \textbf{83.63} & \textbf{7.64} & \textbf{1.03} \\
& Volume regularization & \textbf{90.85} & 83.62 & 7.70 & 1.04 \\
& Moment regularization & 90.84 & 83.60 & 7.67 & \textbf{1.03} \\
& Anatomical relation & 88.98 & 80.65 & 8.23 & 1.27 \\
\midrule
\multicolumn{6}{l}{\textbf{MM-WHS CT, \(64^3\), selected architectural priors}} \\
& Baseline & \textbf{90.85} & \textbf{83.63} & 7.64 & \textbf{1.03} \\
& HM multilayer & 90.60 & 83.23 & 7.78 & 1.06 \\
& 2-Decoder & 90.73 & 83.43 & 7.58 & 1.06 \\
& Cascaded & 90.32 & 82.74 & \textbf{7.55} & 1.08 \\
\midrule
\multicolumn{6}{l}{\textbf{MM-WHS CT, \(128^3\), selected architectural priors}} \\
& Baseline & \textbf{92.05} & \textbf{85.78} & 7.35 & \textbf{0.88} \\
& HM multilayer & 91.80 & 85.38 & 7.28 & 0.90 \\
& 2-Encoder & 92.02 & 85.70 & 7.40 & 0.89 \\
& Cascaded & 92.04 & 85.70 & \textbf{7.26} & 0.89 \\
\midrule
\multicolumn{6}{l}{\textbf{WHS++ CT, \(64^3\), shape-aware losses}} \\
& Baseline & 88.93 & 81.01 & 18.47 & 1.68 \\
& Volume regularization & 89.09 & 81.26 & \textbf{17.91} & 1.63  \\
& Moment regularization & \textbf{89.16} & \textbf{81.47} & 18.31 & \textbf{1.61} \\
& Anatomical relation & 88.66 & 80.67 & 18.13 & 1.70 \\ 
\midrule
\multicolumn{6}{l}{\textbf{WHS++ CT, \(64^3\), selected architectural priors}} \\
& Baseline & \textbf{88.93} & \textbf{81.01} & \textbf{18.47} & \textbf{1.68} \\
& HM multilayer & 88.72 & 80.63  & 20.88 & 1.73 \\
& 2-Encoder & 87.75 & 79.43 & 18.52 & 1.77 \\
& Cascaded & 86.13  & 77.05 & 21.27 & 2.09 \\
\bottomrule
\end{tabular}
\end{table}

At the architectural level, heatmap-guided models remained competitive but did not clearly surpass the reference U-Net. At \(64^3\), the 2-Decoder variant reached \(90.73\%\) Dice and the cascaded model produced the best HD (\(7.55\) mm), suggesting slightly improved boundary refinement, but overall overlap remained below baseline. At \(128^3\), the baseline improved to \(92.05\%\) Dice, while the closest competitors, 2-Encoder and Cascaded, achieved \(92.02\%\) and \(92.04\%\), respectively. Hence, higher resolution improved most models, but not our main finding.

%Evaluation on WHS++ confirmed the same trend. The baseline achieved \(88.93\%\) Dice, compared with \(89.09\%\) for Volume and \(89.16\%\) for Shape moment regularization. These differences were small and inconsistent, and more complex prior-guided architectures did not yield a robust advantage across the experiments. Overall, the main finding was stable across datasets and model families.

Evaluation on WHS++ confirmed the same trend. The baseline achieved \(88.93\%\) Dice, while volume and moment regularization yielded only marginal changes, reaching \(89.09\%\) and \(89.16\%\) Dice, respectively. Moment regularization produced the best Dice, Jaccard, and ASSD, whereas volume regularization achieved the lowest HD (17.91 mm). However, these improvements were small and inconsistent. Architectural prior-based models did not outperform the baseline: HM multilayer and 2-Encoder showed slightly lower overlap, and the cascaded architecture degraded performance across all metrics. Overall, the main finding remained consistent across datasets and model families.

\begin{figure}[t]
\centering
\begin{minipage}[t]{0.19\linewidth}
\centering
\includegraphics[width=\linewidth]{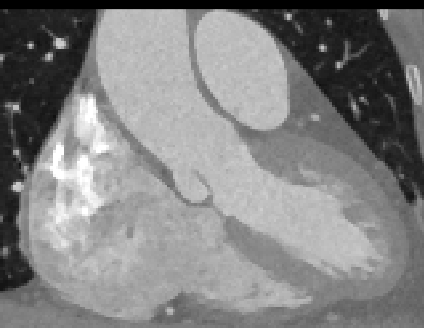}\\
{\scriptsize raw image}
\end{minipage}\hfill
\begin{minipage}[t]{0.19\linewidth}
\centering
\includegraphics[width=\linewidth]{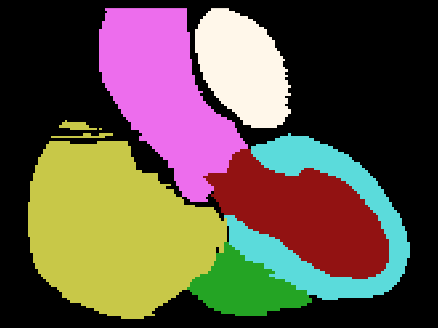}\\
{\scriptsize ground truth}
\end{minipage}\hfill
\begin{minipage}[t]{0.19\linewidth}
\centering
\includegraphics[width=\linewidth]{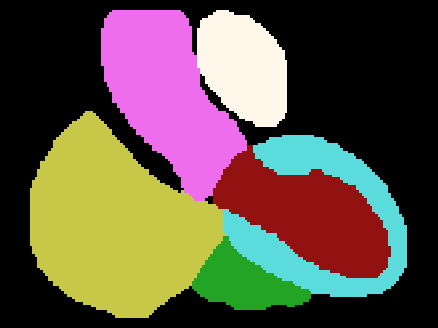}\\
{\scriptsize baseline}
\end{minipage}\hfill
\begin{minipage}[t]{0.19\linewidth}
\centering
\includegraphics[width=\linewidth]{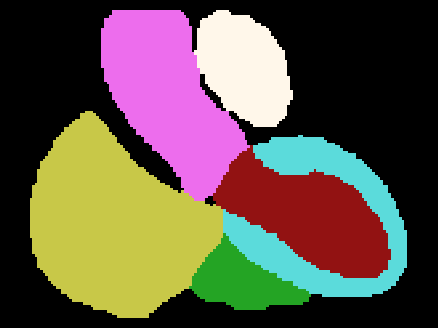}\\
{\scriptsize shape loss}
\end{minipage}\hfill
\begin{minipage}[t]{0.19\linewidth}
\centering
\includegraphics[width=\linewidth]{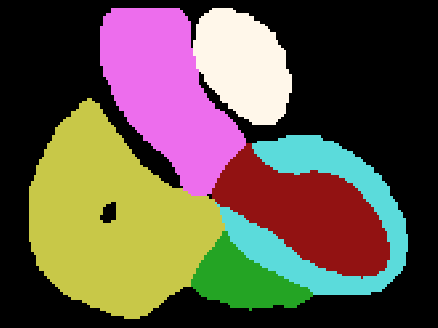}\\
{\scriptsize 2-Decoder}
\end{minipage}
\caption{Representative qualitative comparison on WHS++ (subject 2014, coronal slice 76). The baseline U-Net, the best loss-based prior (Mean-Shape), and the best architecture-based prior (2-Decoder) all capture the global anatomy well. Differences are mainly confined to boundaries and smaller structures.}
\label{fig:qualitative}
\end{figure}

Qualitatively, all models reproduced the overall cardiac configuration and inter-structure arrangement well (Fig.~\ref{fig:qualitative}). The shape-aware variants appeared visually very similar to the baseline, with differences concentrated at boundaries and in thinner structures rather than in gross anatomical localization. This visual pattern is consistent with the quantitative results: the baseline already learned strong global anatomical regularities, leaving limited room for coarse handcrafted priors to add useful information.

\section{Discussion and Conclusions}
The central result of this study is that, surprisingly, explicit handcrafted shape priors did not consistently outperform a strong 3D U-Net baseline for whole-heart CT segmentation, which is performing in-line with the winner of the MM-WHS Challenge (see~\cite{Payer2018-ma,Zhuang2019-uj}) as well as the participants at the Challenge associated with WHS++~\cite{Thaler2025-xb}. This is a relevant negative result. It suggests that on MM-WHS and WHS++, the baseline model already learns substantial implicit anatomical regularities directly from the image data, and that coarse constraints such as expected volumes, low-order moments, centroid relations, or average label distribution heatmaps add little information beyond that baseline.

Several factors likely explain this outcome. First, the tested priors describe anatomy only at a coarse level and cannot capture the complex nonlinear variability of cardiac shape. Second, the remaining errors are mainly boundary-related, whereas the priors regularize global structure more strongly than local boundary detail. Third, because baseline performance is already high, measurable improvements are inherently limited. The experiments also show that greater architectural complexity does not automatically improve segmentation: models such as 2-Encoder and Cascaded remained competitive, and Cascaded slightly improved HD, but none clearly surpassed the simpler baseline in overall Dice or boundary overlap.

In summary, this work provides a focused evaluation of explicit shape priors in deep learning-based whole-heart segmentation and shows that simple handcrafted priors are insufficient to reliably improve a strong 3D U-Net. Future work should therefore move toward more expressive learned anatomical priors, such as generative diffusion-based~\cite{Ho2020-sk} or flow matching-based~\cite{Hadzic2025-rm} models trained on segmentation masks, which may better represent the distribution of plausible cardiac anatomy.

\begin{ack}

This research was funded in whole or in part by the Austrian Science Fund (FWF) 10.55776/PAT1748423.

%Use unnumbered first-level headings for the acknowledgments. All acknowledgments are put at the end of the paper before the list of references. Moreover, you are required to declare
%funding (financial activities supporting the submitted work) and competing interests (related financial activities outside the submitted work).
%More information on this disclosure can be found at \url{https://neurips.cc/Conferences/2023/PaperInformation/FundingDisclosure}.

%Do {\bf not} include this section in the anonymized submission, only in the final paper. You can use the \texttt{ack} environment provided in the style file to automatically hide this section in the anonymized submission.
\end{ack}

\bibliographystyle{unsrt}
\bibliography{paper}
%%%%%%%%%%%%%%%%%%%%%%%%%%%%%%%%%%%%%%%%%%%%%%%%%%%%%%%%%%%%

\end{document}